\documentclass[10pt]{article}
\usepackage{psfig}
\usepackage{emulateapj}
\usepackage{apjfonts}
\usepackage{timesfonts}

\def\mes{M{\'e}sz{\'a}ros}

\begin{document}

\title{Soft X-ray emission lines in the early afterglow of 
gamma-ray bursts}

\author{Davide Lazzati, Enrico Ramirez-Ruiz and Martin J. Rees} 
\affil{Institute of Astronomy, University of Cambridge,
Madingley Road, Cambridge CB3 0HA, England \\ 
{\tt lazzati@ast.cam.ac.uk}}

\begin{abstract}
We compute the luminosity of $K_\alpha$ emission lines produced by
astrophysically abundant elements in the soft X-ray spectra of the
early afterglow of gamma-ray bursts. We find that the detection of
these lines can be a diagnostic for the geometrical set-up of the
reprocessing material.  In particular we can distinguish between a
``geometry dominated'' model, in which the line emission is coming
from an extended region and its duration arises from light-travel-time
effects and an ``engine dominated'' model, where the line emitting gas
is in a smaller region, irradiated for a longer period. These lines
therefore offer clues to the dynamics and time-scale of the explosion
leading to a gamma-ray burst.
\end{abstract}

\section{Introduction}

Narrow emission features superimposed on the early X-ray afterglow of
gamma-ray bursts (GRBs) seem to be common [Piro et al. 1999, Antonelli
et al. 2000, Piro et al. 2000 (hereafter P00), Reeves et al. 2002
(hereafter R02)], even though a single strong evidence of their
significance is still lacking. These features are observed starting
from several hours after the burst (due to instrumental limitations)
up to a couple of days. Some have signs of variability (e.g. R02),
others are constant for the total duration of the observation
(Antonelli et al. 2000; P00).  Understanding how and where they are
produced is fundamental for identifying the nature of the burst
progenitor. In fact the line, being observed at its rest wavelength,
has to be produced through reprocessing of the burst radiation, and
carries therefore information about the geometry and structure of the
material surrounding the burst (Lazzati, Campana \& Ghisellini
1999). It is commonly accepted that the presence of metal lines itself
strongly favors hypernova models (Woosley 1993; Paczynski 1998) and
can be used to rule out compact mergers (Eichler et al. 1989) as burst
progenitors. Nevertheless, whether the lines can be produced in a
standard hypernova scenario or require an explosive event or supernova
that occurred prior to the GRB (see e.g. Vietri \& Stella 1998) is
still a matter of open debate.

We define as Engine Dominated (ED) the models in which the lines are
produced in a standard -- single step -- hypernova explosion. In these
models, the line is created by reprocessing from material very close
to the explosion site ($R\sim10^{13}$~cm). The ionizing continuum in
this case can not be attributed to the burst or afterglow itself,
since the photons are radiated at larger distances. The ionizing
continuum is then believed to be provided by a long lasting engine
(Rees \& \mes~2000) or by magnetic energy stored in a plasma bubble
emerging after the jet has crossed the stellar progenitor (\mes~\&
Rees 2001). In ED models the duration of line emission is related to
the duration of the the ionizing continuum.

We define as Geometry Dominated (GD) the models in which the
reprocessing material is located at a large enough distance, $R$, to
be illuminated by the burst and afterglow photons. In these models the
duration of the line emission is set by the size of the
reprocessor. This reprocessing material has to be compact and metal
enriched, similar to a supernova remnant (SNR), as naturally predicted
in the supranova model (Vietri \& Stella 1998).  In both ED and GD
models the line is supposed to be produced by reflection off a slab of
optically thick material (see also B\"ottcher 2000 and B\"ottcher \&
Fryer 2001 for an alternative GD model).

In this letter we compute the luminosity of $K_\alpha$ lines from the
elements O, Ne, Mg, Si, S, Ar and Ca (which we will refer to as
``light elements'' hereafter), and compare them to the iron one. We
find that the detection of soft X-ray ($\sim[0.5-5]$~keV) features,
possibly made in the afterglow of GRB~011211 (R02) strongly favors GD
scenarios for the production of metal lines in GRB afterglows.  The
letter is organized as follows. In \S 2 we compute the line
luminosities in reflection from an {\it a priori} point of view. We
compute the line flux in GD and ED scenario in \S 3 and we discuss the
comparison of predictions with observational constraints in \S 4.

\section{Line luminosities}

The luminosity of the lines has been computed according to the equation:
\begin{equation}
L_X={{n_X\,\epsilon_X\,S\,\lambda}\over{t_{\rm em}}}
\label{eq:lum}
\end{equation}
where the subscript $_X$ indicates the particular ion, $n_X$ is the
ion density, $\epsilon_X$ is the $K_\alpha$ emission line frequency
$\epsilon_X=h\nu_X$ of the H-like ion, $S$ is the emitting surface,
$\lambda$ is the depth of the emitting layer and $t_{\rm em}$ is the
time scale of emission of a photon by a single ion.

The time scale of photon emission $t_{\rm em}$ can be set either by the
ionization time scale corrected for the Auger effect (if the ionization
parameter is low) or by the recombination time scale, for highly
ionized plasma. Here, we compute the emission time as
$\max(t_{\rm{rec}},\,t_{\rm{ion}}/Au_{\{X,Z\}})$, where
$Au_{\{X,Z\}}$ is the photoelectric yield for the element $X$ with
charge $Z$ (Kaastra \& Mewe 1993). The ionization time scale
$t_{\rm{ion}}$ is computed as (Lazzati et al. 2001b):
\begin{equation}
t_{\rm ion} = (3+\alpha) {{4\pi\,R^2\,h\,\nu_X^\alpha}\over{L_0\,\sigma_X}}
\label{eq:tion}
\end{equation}
where $L(\nu)=L_0\,\nu^{-\alpha}$ is the ionizing spectrum (note that
$L_0$ does not have the dimensions of a luminosity), $\sigma_X$ the
threshold cross section of the element $X$ (atomic data are taken from
Verner \& Yakovlev 1995) and $R$ the distance of the reprocessing
material from the ionizing flux source. The recombination time scale
$t_{\rm{rec}}$ has been computed with the analytic approximations and
tables of Verner \& Ferland (1996). A constant temperature
$T=4\times10^7$~K has been assumed. Both the ionization and
recombination time scales are computed for the H-like ion.  The charge
of the atom is then computed as\footnote{The exponent $2/5$ has been
derived by interpolation of the numerical recombination rates of Shull
\& van Steenberg (1982).} $Z_X=(t_{\rm{ion}}/t_{\rm{rec}})^{2/5}$.

The depth of the emitting layer $\lambda$ is assumed to be the depth
that the ionizing photons can reach and/or a line photon can escape
without a shift in frequency. We therefore compute it as the
minimum between the plasma opacity at resonance for the ion $X$ and
the maximum Thompson depth a photon can travel before being shifted by
Thomson scattering. We therefore have
$\lambda=\min(\lambda_T,\,\lambda_X)$, where:
\begin{equation}
\lambda_X = \left[ \sum_Y{{n_Y\,t_{\{\rm{ion, Y}\}}\,\sigma_Y(\nu_X)}
\over{t_{\{\rm{ion, Y}\}}+t_{\{\rm{rec, Y}\}}}} \right]^{-1}
\label{eq:pippa}
\end{equation}
and
\begin{equation}
\lambda_T = {{\sqrt{N_{\rm{sc}}(T)}}\over{n_e\,\sigma_T}}
\label{eq:lamt}
\end{equation}
where the sum in Eq.~\ref{eq:pippa} is extended over the
astrophysically relevant elements H, He, C, N, O, Ne, Mg, Si, S, Ar,
Ca, Fe (elemental abundances taken from Anders \& Grevesse 1989), and
$\sigma_Y(\nu_X)$ is the cross section of element $Y$ computed at the
threshold frequency of element $X$. In the numerator of
Eq.~\ref{eq:lamt}, $N_{\rm{sc}}(T)$ is the maximum number of
scatterings a photon can make before the line is smeared or shifted.
If $\epsilon\sim{kT}$, line broadening is more important than
shifting, and we have
$\sigma_\epsilon/\epsilon\sim\sqrt{kT/(m_e\,c^2)}$ for a single
scattering. Requiring a limit of $\sigma_\epsilon/\epsilon\le1/3$ in
order for the line to be detectable as a narrow feature, we have
$\tau_{\max}=\sqrt{N_{\rm{sc}}}\sim[m_e\,c^2/(9\,k\,T)]^{1/2}\approx4$. Note
however that, for the considered abundant elements, Eq.~\ref{eq:lamt}
is relevant only for very high ionization parameters (i.e. in the
decreasing tails of Fig.~\ref{fig:lr}) and therefore its uncertainty
is not necessarily reflected in the maximum efficiency of line
reprocessing.

In Fig.~\ref{fig:lr} and~\ref{fig:fer} (lower panel) we show the
results of our calculations for O, Ne, Mg, Si, S, Ar, Ca and Fe as a
function of the ionization parameter $\xi=4\pi\,F_{\rm{[1-10]}}/n$,
where $F_{\rm{[1-10]}}$ is the ionizing flux in the energy range
[1-10]~keV and $n$ the particle density. The figures show that, in
optimal conditions, $\sim1\%$ of the ionizing flux is reprocessed in
each line. We also show that increasing the metallicity does not
increase this fraction, but helps keep it constant for high ionization
parameters. The accuracy of this line strength estimates have been
checked against the results of numerical simulations for Mg, Si and Fe
(Ross 1979; Ross \& Fabian 1983; Ballantyne, Fabian \& Ross 2002). We
find that our results are in agreement within a factor $\sim 2$ with
line luminosities obtained by fitting Gaussian profiles to the
numerical spectra. The largest deviations are observed close to the
minimum value of the efficiency (in the strong Auger regime), where
our semi-analytical approach yields larger luminosities with respect
to the full numerical treatment. This comparison has been performed
with Mg, Si and Fe lines, since the numerical code does not include S,
Ar and Ca.

\section{GRB models} 

In order to apply the results of the previous section to various GRB
models, we need to transform the ratios shown in Fig.~\ref{fig:lr}
and~\ref{fig:fer} into observed fluxes and/or luminosities. Even
though in spectroscopy it is customary to compute the line equivalent
width, in the case of GRBs it is assumed that the observed continuum
is not related to the ionizing continuum responsible for producing the
lines (which can be the afterglow itself at earlier times in GD models
or a completely different component in ED models), and so it is more
useful to directly compare the line fluxes (Lazzati et al. 1999).

\subsection{Geometry dominated models}

In the case of GD models, the observed line flux has to be corrected
by a geometric factor $\zeta$. This factor takes into account that
while the emitting material is illuminated for a time $t_{\rm{ill}}$,
the line is observed, at infinity, for a time
$t_{\rm{obs}}\sim{R}_{\rm{SNR}}/c$.  A precise computation of
$\zeta\equiv{t}_{\rm{ill}}/t_{\rm{obs}}$ is complex, since the
reprocessing material is illuminated by the burst and early afterglow
radiation, and it is likely that the properties of the fireball
towards the reprocessing material are different from those along the
line of sight (Rossi, Lazzati \& Rees 2002).  If one assumes that the
fireball decelerates inside the remnant and that the density of the
external medium in the immediate vicinity of the progenitor scales
with the remnant radius\footnote{Note that the density is not supposed
to scale with the radius. We consider a uniform density inside the
remnant, scaling it with the remnant radius in order to have a
constant total mass inside it.} as
$n=10^8\,n_8\,(R_{\rm{SNR}}/10^{15})^{-3}$~cm$^{-3}$, the time the
fireball takes to overtake the remnant is proportional to the radius
of the remnant itself, and:
\begin{equation}
\zeta \sim {{c\,(t_{\rm{GRB}}+t_{\rm{ill}})}\over{R_{\rm{SNR}}}} \approx
{{c\,t_{\rm{ill}}}\over{R_{\rm{SNR}}}}\approx
0.05 {{n_8}\over{E_{51}}}
\end{equation}
where $t_{\rm{GRB}}$ is the burst duration, and $E_{51}$ is the
isotropic equivalent energy output of the burst in units of
$10^{51}$~erg in the direction of the reprocessing material. We have
used Eq.~15 of Panaitescu \& Kumar (2000) to evaluate the numerical
factor.

On the other hand, in GD models the ionizing continuum is not directly
observed, and can then be assumed to be much larger than the afterglow
luminosity at the time the line is detected. The ratio
$\eta_{\rm{cont}}$ of reflected to incident continuum depends on the
ionization parameter, ranging from $<0.1$ for $\xi\le10^2$ to $\sim 1$
for $\xi\ge10^5$ (see the upper panel of Fig.~\ref{fig:fer}). For
reasonable ionization parameters, one can then assume an ionizing
luminosity
$L_{\rm{ion}}\le{L}_{\rm{After}}/(\zeta\,\eta_{\rm{cont}})\sim2\times10^{48}$
erg~s$^{-1}$, where $L_{\rm{After}}$ is the afterglow [1-10] keV
luminosity at the time at which the lines are observed. The apparent
contradiction of having an ionizing continuum brighter than the
afterglow but not observed in the afterglow itself, is solved if we
consider the geometry of the system (see e.g. panels a and b of Fig. 2
in Vietri et al. 2001). In fact, the fireball emission towards the
reprocessing material is not necessarily related to the one in the
direction of the observer. Moreover, line photons produced at time $t$
are observed at a later time due to the longer travel thay have to
make. At this later time the afterglow emission has decreased.

Finally, we must take into account that while the ionizing continuum
may be geometrically beamed, the line emission is nearly isotropic, so
that the observed flux must be corrected by a factor
$\Omega/4\pi\sim0.25$, where the numerical factor has been computed
for a jet opening angle of $\theta=40^\circ$. This opening angle is
one order of magnitude larger than the opening angle measured in
GRB~991216 (Frail et al. 2001). Since
$\Omega/4\pi=1-\cos\theta_j\propto\theta_j^2$, a ten times smaller
opening angle would make the lines two orders of magnitude fainter and
thus undetectable. We here assume this larger angle without discussing
its implications, referring the reader to the thorough discussion
presented in Ghisellini et al. (2002).

The resulting line luminosity of the element $X$ is then given by:
\begin{equation}
L_X={{\Omega}\over{4\pi}}\zeta\,L_{\rm{ion}}\,\eta_X\sim
0.25{{\eta_X}\over{\eta_{\rm{cont}}}}\,L_{\rm{After}}
\label{eq:gd}
\end{equation}
where $\eta_X$ is the luminosity ratio plotted in
Fig.~\ref{fig:lr}. Note that the geometric factor $\zeta$ is not
relevant in this expression.  For low ionization parameters $\xi<10^3$
one has $\eta_{\rm{cont}}\sim0.1$ and $\eta_X\sim10^{-2}$, while for
iron one obtains $\eta_{\rm{Fe}}\sim10^{-3}$ since its efficiency is
strongly decreased by Auger auto ionization (see also Ross, Fabian \&
Brandt 1996).  Light elements can reprocess into $K_\alpha$ line
emission a sizable fraction of the continuum, yielding luminosities of
the order of $L_X\sim2.5\times10^{44}$~erg~s$^{-1}$
($F_X\sim5\times10^{-14}$~erg~cm$^{-2}$~s$^{-1}$ at $z\sim1$).  In
contrast, iron $K_\alpha$ emission will be one order of magnitude
fainter, and therefore undetectable with present instrumentation. For
larger ionization parameters $10^3<\xi<10^4$, iron emission is
efficient, while light elements are too ionized to emit line photons.
A bright iron line should then be detectable in absence of light
element features (see also Ballantyne \& Ramirez-Ruiz 2001).

\subsection{Engine dominated models}

Evaluating the line luminosity in ED models is simpler and less
affected by uncertainties. In fact, there is no geometric dilution of
the photons and the ionizing continuum is likely to be smaller than the
afterglow luminosity at the time of the line observation. Only the
beaming factor $\Omega/4\pi$ has to be taken into account, yielding a
luminosity:
\begin{equation}
L_X={{\Omega}\over{4\pi}}\,L_{\rm{After}}\,\eta_X
\sim0.25\,L_{\rm{After}}\,\eta_X
\label{eq:ed}
\end{equation}
In ED models, then, the line luminosity is a factor
$\eta_{\rm{cont}}$ less intense than in GD models.  Lines from
light elements (Mg, Si, S, Ar and Ca), should be very difficult to
detect since  $L_{\{\rm{Ar,Ca}\}}\le10^{43}$~erg~s$^{-1}$
($F_X\sim10^{-15}$~erg~cm$^{-2}$~s$^{-1}$ at $z\sim1$).  Iron lines
should instead be detectable, in particular if the ionization is not
too high and the spectral slope is flat (Ballantyne et al. 2002).

\section{Summary and discussion}

The presence of emission features in the early X-ray afterglows of
GRBs provides important clues for identifying the burst progenitor. We
computed the luminosity of $K_\alpha$ emission features from
astrophysically relevant elements (Mg, Si, S, Ar and Ca) in reflection
models and applied the results to the detectability of those features
in the early afterglows of GRBs. We find that the line luminosity of
these elements can be a strong indicator of the geometry of the
reprocessing material.  In particular, light element lines are
detectable only if the ionizing continuum is not directly
observable. This geometrical set-up is naturally accounted for in GD
models, in which the line photons reach the observer after being
reflected by material away from the line of sight. In ED models, the
ionizing continuum and the line are observed simultaneously and for
this reason the strength of light elements lines is greatly diluted,
making them undetectable.

Recently, a possible detection of $K_\alpha$ emission lines from light
elements in the early afterglow of GRB~011211 has been claimed
(R02). Reeves et al. claim that their spectrum is inconsistent with a
reflection model.  We should however be aware that the reflection
model used to fit the data does not include emission from S, Ar and Ca
(Ross \& Fabian 1983). In addition, they fit a pure reflection model,
without including the power-law external shock emission. This
power-law component must however dominate the continuum, since the
broad band lightcurve decreases as a power-law in time (R02).  Let us
now consider the $K_\alpha$ line of sulphur at $2.6$~keV. It is
detected at a flux level of
$(1\pm0.3)\times10^{-14}$~erg~cm$^{-2}$~s$^{-1}$, when the [0.2-10]
keV continuum has a flux of
$\sim2\times10^{-13}$~erg~cm$^{-2}$~s$^{-1}$. This flux ratio
$\eta_S=0.05\pm0.015$ can be compared to the lower left panel of
Fig.~\ref{fig:lr} and Eq.~\ref{eq:ed}. Even if the beaming correction
approaches unity, the detected flux ratio exceeds the maximum allowed
in an ED model, regardless of both the value of the spectral slope and
ionization parameter. In GD models, however, the luminosity of light
element features can be ten times larger due to the very small
fraction of the continuum that is reflected for small ionization
parameters (see the upper panel of Fig.~\ref{fig:fer}).

Another important particularity of the observation of light element
features in GRB~011211 is the lack of an iron line. If we compare this
observation to that of the iron $K_\alpha$ line in GRB~991216 (P00),
we find that the ratio of iron to sulphur luminosities is
significantly different. The line ratio is
$L_{\rm{S}}/L_{\rm{Fe}}<0.4$ in GRB~991216 and
$L_{\rm{S}}/L_{\rm{Fe}}>2$ in GRB~011211. This difference can be
explained, as a consequence of the different ionization state of the
reprocessing material -- in GRB~011211 the reprocessing material had
an ionization parameter between $30\le\xi\le300$, while in GRB~991216
$10^3<\xi<10^5$. In the framework of GD models, the different
ionization parameter can be interpreted as a different age of the SNR
at the time of the burst explosion. If the remnant itself has a
density that scales with the radius (and hence with age)
$n_{\rm{SNR}}\propto{R}_{\rm{SNR}}^{-3}$, the ionization parameter
should also scale with radius for a constant ionizing
luminosity. Young remnants should then produce light element lines,
possibly in coincidence with weak nickel and/or cobalt lines, while
old remnants should produce high ionization iron lines, and no light
element features (see also Vietri et al. 2001). In addition, the
variability time scale of compact remnants should be smaller. All
these predictions are confirmed by observations -- the features
observed in GRB~011211, in which a marginal evidence of Ni line is
present, are observed to disappear after $\sim4$~h (R02), while
the iron line in GRB~991216 is detected $\sim18$~h after the burst
(P00). Is should also be remarked that these conclusions do not depend
on the assumption of a constant temperature of the reprocessing
material. In fact, should this assumption be wrong or the temperature
be estimated incorrectly, the result would be an horizontal shift of
the lines in Fig.~\ref{fig:lr}. Since we have not discussed in which
conditions a certain value of the ionization parameter is relevant,
this does not affect the generality of our conclusions.

In conclusion, the detection of Mg, Si, S, Ar and Ca lines with
equivalent widths of hundreds of eV in the early afterglow of
GRB~011211 can not be accounted for in any version of ED models.  A
two step explosion is necessary, in order to place the reprocessing
material at a distance large enough to explain the duration of the
line and to let the ionizing continuum fade (hidden by the afterglow
emission) before the lines can be observed. This evidence adds to the
transient absorption feature detected in GRB~990705 (Amati et
al. 2000; Lazzati et al. 2001a) which strongly favors a GD
model. Alternatively, the reflecting material may have been expelled
during the common envelope phase of a binary progenitor system
composed by a compact object and a helium star (B\"ottcher \& Fryer
2001). In this case, all the discussion about GD models hold true, but
a Ni line is not expected since the reprocessing material is not a SN
explosion.

\acknowledgements
We are very grateful to K. Pounds and J. Reeves for useful discussions
on the data of GRB~011211 and to A. Fabian and D. Ballantyne for many
discussions on reflection spectra, and for making their numerical
spectra available to us. We thank Markus B\"ottcher for useful
discussions.

\newpage

\centerline{\psfig{file=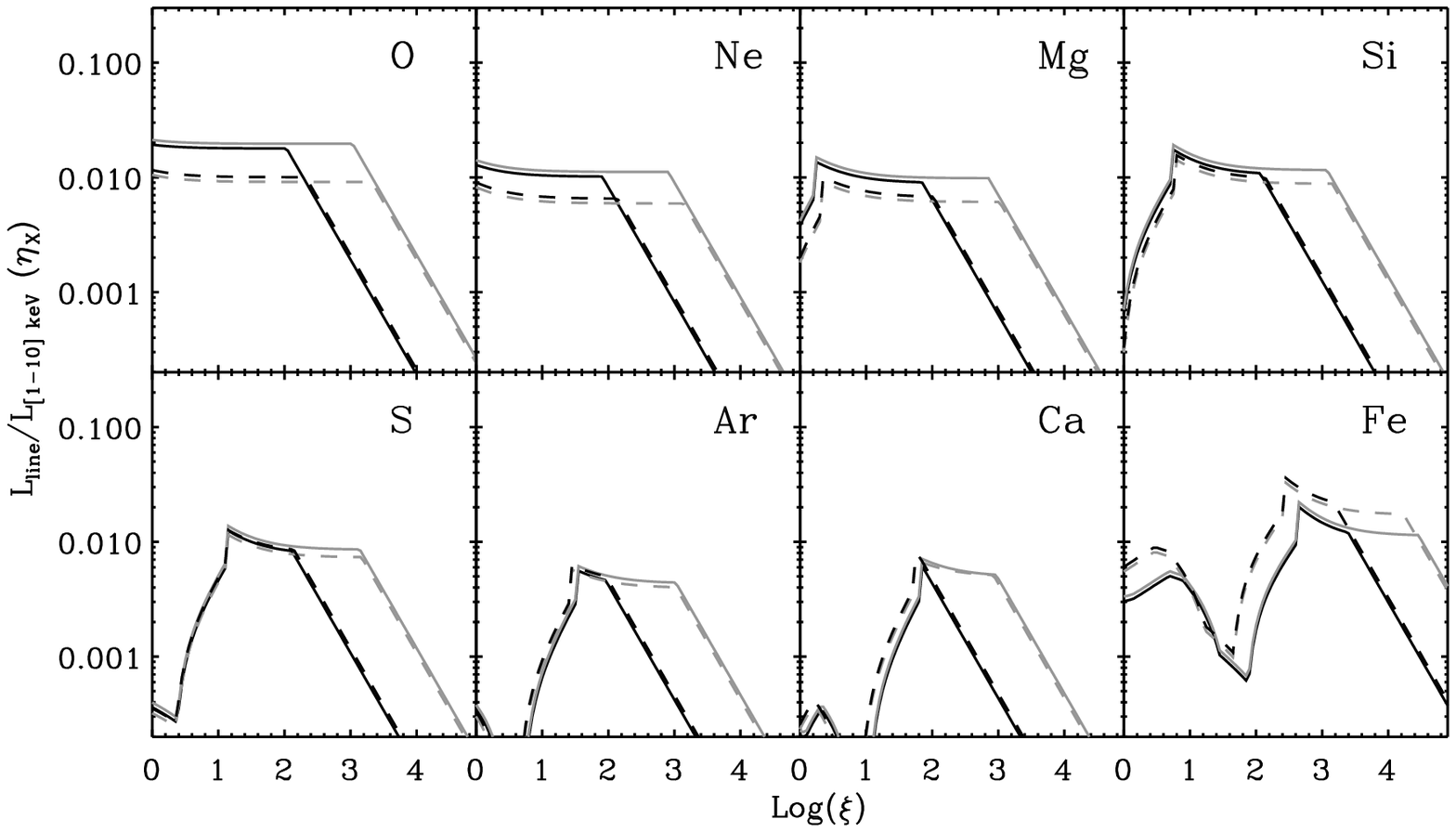,width=0.9\textwidth}}
\figcaption{{Ratio of the $K_\alpha$ line to the [1-10] keV 
continuum luminosities for various astrophysically relevant elements
as a function of the ionization parameter $\xi\equiv
4\pi\,F_{[1-10]}/n$. Solid and dashed lines correspond to a power-law
index $\alpha=1.25$ and $\alpha=0.75$ of the ionizing continuum,
respectively. Black and grey lines correspond to solar and ten times
solar metallicity of the reprocessing material, respectively.}
\label{fig:lr}}

\bigskip
\bigskip
\bigskip

\centerline{\psfig{file=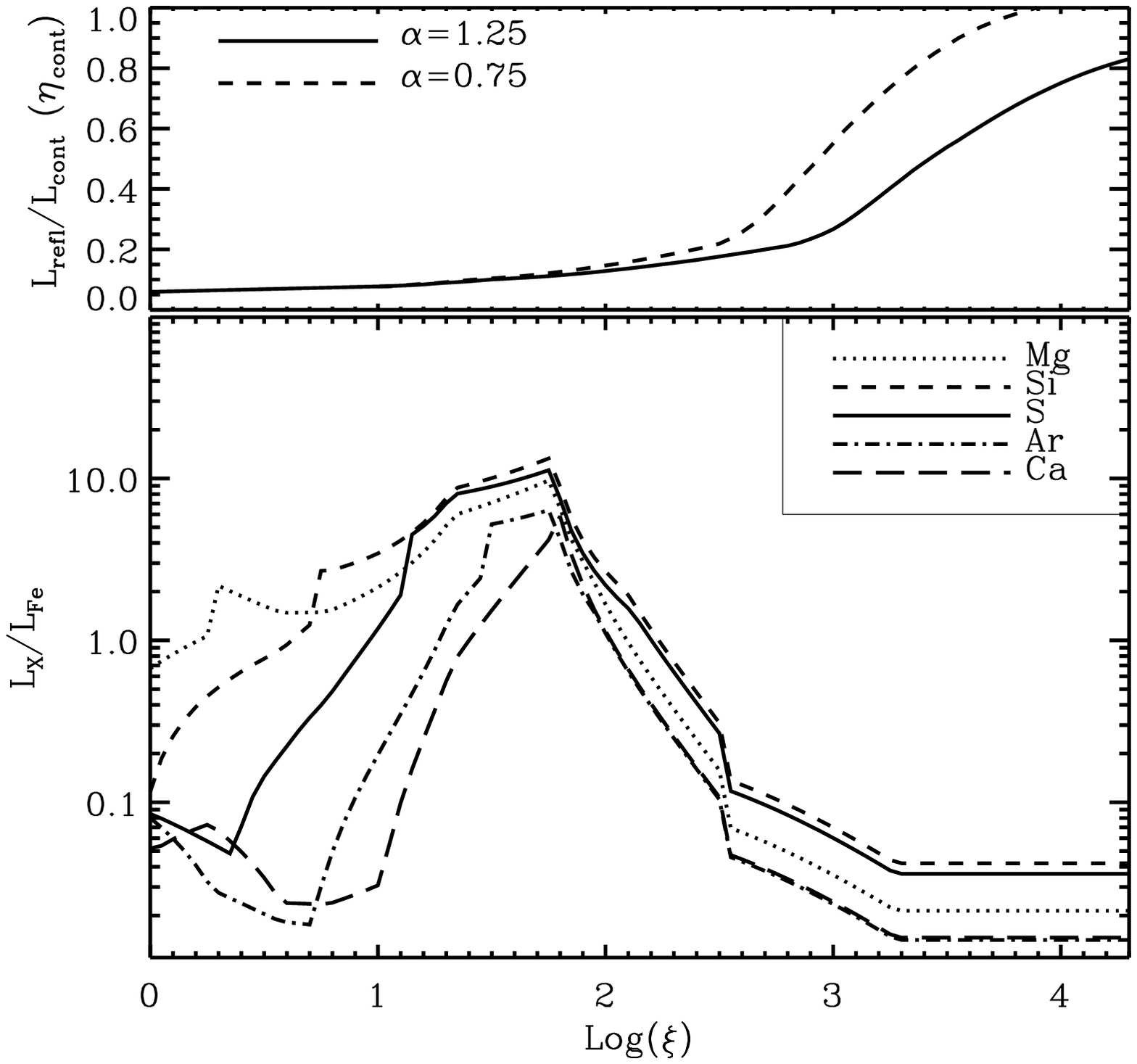,width=0.5\textwidth}}
\figcaption{{Upper panel: ratio of the reflected to incident continuum 
in the [1-10] keV band for a solar metallicity slab. Two incident
spectral slopes are considered, as in Fig.~\ref{fig:lr}.  Lower panel:
ratio of $K_\alpha$ line luminosities of Mg, Si, S, Ar and Ca to the
iron $K_\alpha$ luminosity. Even though the iron line is generally
more intense than all the other lines, there is a range of ionization
parameters $\xi\sim10^2$ for which iron $K_\alpha$ emission is
quenched by Auger auto ionization, and light elements dominate. This
figure is made for a continuum power-law index $\alpha=1.0$ and solar
metallicity. Changing these parameters can be a moderate effect on the
shape of the curves (see Fig.~\ref{fig:lr}).}
\label{fig:fer}}

\end{document}